\documentclass{article}
\def\ba{\begin{array}}
\def\ea{\end{array}}
\def\be{\begin{equation}}
\def\ee{\end{equation}}

\def\a{{\alpha}}
\def\s{{\sigma}}
\def\d{{\rm d}}

\def\o{{\omega}}
\def\O{{\Omega}}
\def\bea{\begin{eqnarray}}
\def\eea{\end{eqnarray}}
\def\b{\beta}

\def\l{\lambda}
\def\L{\Lambda}

\def\b{\beta}
\def\A{{\cal A}}
\def\N{{\cal N}}
\def\p{\emptyset}

\def\t{\tau}

\def\G{\Gamma}
\def\g{\gamma}

\def\rd{{\rm d}}

\begin{document}
\begin{titlepage}
\hfill{}
\vskip 5 mm
\noindent{ \Large \bf
         Multispecies reaction--diffusion systems}

\vskip 1 cm
\noindent{A. Aghamohammadi$^{1,3,{\rm a}}$, A. H. Fatollahi$^{2,3,{\rm b}}$,
          M. Khorrami$^{2,3,{\rm c}}$, A. Shariati$^{2,3,{\rm d}}$ }
\vskip 5 mm
{\it
  \noindent{ $^1$ Department of Physics, Alzahra University,
             Tehran 19834, Iran. }

  \noindent{ $^2$ Institute for Advanced Studies in Basic Sciences,
             P.O.Box 159, Gava Zang, Zanjan 45195, Iran. }

  \noindent{ $^3$ Institute for Studies in Theoretical Physics and
            Mathematics, P.O.Box  5531, Tehran 19395, Iran. } }
\\[\baselineskip]  $^{\rm a}$ mohamadi@theory.ipm.ac.ir
\\  $^{\rm b}$ fath@iasbs.ac.ir
\\  $^{\rm c}$ mamwad@iasbs.ac.ir
\\  $^{\rm d}$ shahram@iasbs.ac.ir

\vskip 1 cm


\vskip 1cm

\begin{abstract}
Multispecies reaction--diffusion systems, for which
the time evolution equation of correlation functions become a closed set,
are considered.
A formal solution for the average densities is found. Some
special interactions and the exact time dependence of the average densities
in these cases are also studied.
For the general case, the large time behaviour of the average densities
has also been obtained.
\end{abstract}
\vskip 10 mm
\end{titlepage}
\section {Introduction}
In recent years, reaction--diffusion systems have been studied by many
people. As mean field techniques, generally do not give correct results
for low dimensional systems,
people are motivated to study stochastic models in low dimensions.
Moreover, solving one dimensional systems should in principle
be easier.
Exact results for some models in a one--dimensional lattice have
been obtained, for example in [1--10].

Different methods have been used to study these models, including analytical
and asymptotic methods, mean field methods, and large-scale numerical
methods. Systems with more than one species have also been studied [11--23].
Most of the arguments are based on simulation results. There are, however,
some exact results as well ( \cite{VK,AA,FJ} for example).

In \cite{GS}, a 10--parameter family of stochastic models has been studied.
In these models, the $k$--point  equal time correlation functions
$\langle  n_i n_j\cdots n_k\rangle $ satisfy linear differential equations
involving no higher--order correlations.
These linear equations for the average density $\langle  n_i\rangle $ has
been solved. But these set of equations can not be solved easily for
higher order correlation functions.
We have generalized the same idea to multi-species models. We have considered
general reaction diffusion processes of multi-species in one dimension with
two-site interaction. We have obtained the conditions the Hamiltonian
should satisfy in order to give rise to closed set of time evolution
equation for correlation functions. The set of equations for average
densities can be written in terms of four matrices. The time evolution
equation for more-point functions, besides these four matrices, generally
depend explicitly on the elements of the Hamiltonian , and generally can
not be solved easily.
These matrices are not determined uniquely from the Hamiltonian:
there is a kind of gauge transformation one can apply on them which
of course, does not change the evolution equation. A formal solution
for average densities of different species is found.
For some special choices  of the four matrices we also give the explicit
form of interactions and the exact time dependence of average
densities. At the end, we study the large time behaviour of the average
densities of different species for the general case.

\section{A brief review of linear stochastic systems}
To fix the notation used in this article, here we briefly review
the already well known formalism of linear stochastic systems.
The master equation for $P(\s ,t)$ is
\be
{\partial\over\partial t} P(\s ,t)= \sum_{\t \ne \s}\big[ \o (\t\to \s )
P(\t ,t)-\o (\s \to \t )P(\s ,t) \big] ,
\ee
where $ \o (\t\to \s )$ is the transition rate from the configuration
$\t$ to $ \s$. Introducing the state
vector
\be
\vert P \rangle  = \sum_{\s} P(\s ,t ) \vert \s \rangle ,
\ee
where the summation runs over all possible states of the system, one can
write the above equation in the form
\be \label{sh}
{\d \over \d t} \vert P\rangle ={\cal H} \vert P\rangle ,
\ee
where the matrix elements of ${\cal H}$ are
\bea
\langle   \s \vert {\cal H}\vert \t \rangle = \o ( \t \to \s ),
 \qquad \t \ne \s,\cr
\langle   \s \vert {\cal H}\vert \s \rangle = -\sum_{\t\ne \s}\o
( \s \to \t ).
\eea
The basis $\{\langle  \s \vert\}$ is dual to $\{ \vert\s \rangle \}$,
that is
\be
\langle   \s \vert  \t \rangle =\delta_{\s  \t}.
\ee
The operator  ${\cal H}$ is called a Hamiltonian, and it is not
necessarily hermitian. Conservation of probability,
\be
\sum_{\s} P(\s ,t)=1,
\ee
shows that
\be
\langle  S \vert {\cal H}=0,
\ee
where
\be
\langle   S \vert =\sum_{\b} \langle   \b\vert .
\ee
So, the sum of each column of ${\cal H}$, as a matrix,  should be zero.
As $\langle S \vert$ is a left eigenvector of ${\cal H}$ with zero
eigenvalue,
${\cal H}$ has at least one right eigenvector with zero eigenvalue. This
state corresponds to the steady state distribution of the system and it
does
not evolve in time. If the zero eigenvalue is degenerate, the steady state
is not unique. The transition rates are non--negative, so the off--diagonal
elements of the matrix ${\cal H}$  are non--negative. Therefore, if a matrix
${\cal H}$ has the following properties,
\be\ba{l}
\langle   S \vert {\cal H}=0,\cr
\langle   \s \vert {\cal H}\vert \t \rangle \geq 0,
\ea \ee
then it can be considered as the generator of a stochastic
process. It can be proved that the real part of the eigenvalues of any
matrix with the above conditions is less than or equal to zero.

The dynamics of the state vectors (\ref{sh}) is given by
\be
\vert P(t)\rangle = \exp (t{\cal H}) \vert P(0)\rangle ,
\ee
and the expectation value of an observable ${\cal O}$ is
\be
\langle   {\cal O}\rangle (t)=\sum_{\s } {\cal O}(\s ) P(\s ,t)=\langle
S \vert {\cal O}
\exp (t{\cal H}) \vert P(0)\rangle .
\ee
\section{Models leading to closed set of evolution equations}
The models which we address are multispecies reaction--diffusion models.
That is, each site is a vacancy or has one particle. There are several kinds
of particles, but at any time at most one kind can be present at each site.
Suppose the interaction is between nearest neighbors, and the system is
translationally invariant.
\be
{\cal H}=\sum_{i=1}^L H_{i,i+1}.
\ee
The number of sites is $L$ and the number of possible states
in  a site is $N$; different states of each site are denoted by
$A_{\a}, \quad \a =1,\cdots N$, where one of the states is vacancy.
Introducing $n^{\a}_i$ as the number
operator of $A_{\a}$ particle in the site $i$, we have
\be \label{n1}
\sum_{\a =1}^N  n^{\a}_i =1.
\ee
The average number density of the particle $A_{\a }$ in the site $i$
at the time $t$ is
\be
\langle n^{\a}_i \rangle = \langle  S \vert n^{\a}_i \vert P(t)\rangle
\ee
where $ \vert P(t)\rangle :=\exp (t {\cal H})\ \vert P(0)\rangle $
represents the state of the system at the time $t$,
\be
\langle   S\vert =\underbrace{\langle   s\vert \otimes \cdots \otimes
\langle  s\vert}_{L},
\ee
and
\be
\langle   s\vert :=\underbrace{(1\ 1\ \cdots 1)}_{N}.
\ee
So, the time evolution of $\langle n^{\a}_i \rangle $ is given by
\be\label{dn}
{\d \over \d t}\langle n^{\a}_i \rangle = \langle  S \vert n^{\a}_i
{\cal H}\vert P(t)\rangle .
\ee
The only terms of the Hamiltonian ${\cal H}$ which are relevant in the
above equation are $H_{i,i+1}$ and $H_{i-1,i}$. The result of acting any matrix
$Q$ on the ket $\langle   s\vert $ is equivalent to acting the diagonal matrix
$\tilde Q$ on the same ket, provided each diagonal element of the matrix
$\tilde Q$ is the sum of all elements of the corresponding column in the
matrix $Q$.
So, the action of $(1\otimes n^{\a})H$ and $(n^{\a}\otimes 1)H$  on
$\langle   s\vert \otimes \langle   s\vert $ are equivalent to the action
of two diagonal matrices on $\langle   s\vert \otimes \langle   s\vert $.
We use the notation $\sim $, for the equivalent action on
$\langle   s\vert \otimes \langle   s\vert $.

\bea
(1\otimes n^{\a})H &\sim &\sum_{\b \g }\A ^{\a}_{\b \g }\ \ n^{\b}
\otimes n^{\g}\cr
(n^{\a}\otimes 1)H &\sim &\sum_{\b \g }{\bar \A}^{\a}_{\b \g }\ \ n^{\b}
\otimes n^{\g}.
\eea
where $\A^{\a} _{\b \g }$ and $\bar \A^{\a} _{\b \g } $ are as the following
\bea
\A^{\a} _{\b \g }&:=&\sum_{\l} H^{\l \a }_{\b \g}\cr
\bar \A^{\a}_{\b \g }&:=&\sum_{\l} H^{\a \l}_{\b \g}.
\eea
Then, equation (\ref{dn}) takes the following form
\be \label{ndot}
\langle \dot n^{\a}_i \rangle = \sum_{\b \g } \A^{\a}_{\b \g }
\langle  n^{\b}_{i-1}n^{\g}_i \rangle +\bar \A^\a_{\b \g }
\langle  n^{\b}_{i}n^{\g}_{i+1} \rangle .
\ee
Generally, in the time evolution equation of $\langle n^{\a} \rangle $
the two--point functions $\langle n^{\b} n^{\g} \rangle $ appear.
Using (\ref{n1}), one can see that iff $\A$ and
$\bar \A$ satisfy the following equations, then the right hand side of the
(\ref{ndot}) can be expressed in terms of only one--point functions.
\bea \label{AAA}
\A^\a_{\b \g }+\A^\a_{NN}-\A^\a_{N \g }-\A^\a_{\b N}&=&0,\cr
\bar \A^\a_{\b \g }+\bar \A^\a_{NN}-\bar \A^\a_{N \g }-
\bar \A^\a_{\b N}&=&0.
\eea
These equations give $2(N-1)^3$ constraints on the Hamiltonian, so
adding the condition of stochasticity of $H$, we have $2(N-1)^3+N^2$
relations between  the elements of $H$.
The constraints (\ref{AAA}) mean

\bea \label{sbc}
\A^\a_{\b \g }&=&C^\a_{\b }-B^\a_{\g }\cr
\bar \A^\a_{\b \g }&=&-\bar B^\a_{\b }+\bar D^\a_{\g }.
\eea
So, (\ref{dn}) takes the form
\be \label{ndot2}
\langle \dot n^{\a}_i \rangle = \sum_{\b =1}^N \big[ -(B^{\a}_\b +
\bar B^\a_\b )
\langle  n^{\b}_{i}\rangle +C^\a_\b\langle  n^{\b}_{i-1}\rangle
+\bar D^\a_\b\langle  n^\b_{i+1}\rangle \big] .
\ee
In the simplest case, the one-species, each site is vacant or occupied by
only one kind of particles. Then, the matrices $B$, $C$, $\bar B$, and
$\bar D$ are two-dimensional. Using (\ref{n1}), The equation for
$\langle \dot n^1_i \rangle$  is
\bea\label{2D}
\langle \dot n^1_i \rangle =&& \big( -B^1_1-\bar B^1_1+B^1_2+\bar B^1_2\big)
\langle  n^1_i\rangle +\big( C^1_1-C^1_2 \big) \langle  n^1_{i-1}\rangle\cr
&& +
\big( \bar D^1_1-\bar D^1_2 \big) \langle  n^1_{i+1}\rangle
+\big( - B^1_2-\bar B^1_2+C^1_2+\bar D^1_2 \big) .
\eea
This is a linear difference equation, of the kind obtained \cite{GS}, and
its solution can be expressed in
terms of modified Bessel functions.

The time evolution equation for two-point functions also can be obtained.
\bea
{\d \over \d t} \langle  n^\a_i n^\b_j \rangle &=& \sum_{\g }^N \big[
-(B^\a_\g +\bar B^\a_\g )\langle  n^\g_in^\b_j\rangle +C^\a_\g
\langle  n^\g_{i-1}n^\b_j\rangle +
\bar D^\a_\g\langle  n^\g_{i+1}n^\b_j\rangle \cr &&
-(B^\b_\g +\bar B^\b_\g )\langle  n^\a_in^\g_j\rangle +C^\b_\g
\langle  n^\a_in^\g_{j-1}\rangle\cr
&& +\bar D^\b_\g\langle  n^\a_in^\g_{j+1}\rangle \big],
\qquad \vert i-j \vert >1,
\eea
\bea
{\d \over \d t}\langle  n^\a_i  n^\b_{i+1} \rangle =&& \sum_{\g }^N \big[
-B^\a_\g \langle  n^\g_i  n^\b_{i+1}\rangle
-\bar B^\b_\g \langle  n^\a_i  n^\g_{i+1}\rangle\cr
&&+C^\a_\g \langle  n^\g_{i-1}  n^\b_{i+1}\rangle
+\bar D^\b_\g \langle  n^\a_i n^\g_{i+2} \rangle \big]+
\sum_{\g \l }H^{\a \b}_{\g \l }\langle  n^\g_{i} n^{\l}_{i+1} \rangle .
\eea
For more--point functions, one can deduce similar results.
In fact, it is easy to show that if the evolution equations of
one--point functions are closed, the evolution equation of $n$--point
functions contain only $n$- and less-point functions. However,
generally these set of equations can not be solved easily.

\section{Equivalent Hamiltonians regarding one-point functions, and
gauge transformations}
\noindent Knowing $B$, $C$, $\bar B$, and $\bar D$ does not determine
the Hamiltonian uniquely, but as it is seen from (\ref{ndot2}),
the time evolution of one--point functions depends only on
$B$, $C$, $\bar B$, and $\bar D$.
The two-- and more--point functions depend explicitly on the elements
of $H$. So, different Hamiltonians may give same evolutions for
$\langle n^\a_i \rangle $. Take two Hamiltonians $H$ and $H'$. Defining
\be
R:=H-H',
\ee
if
\be
\sum_{\a} R^{\a \b}_{\g \l }=\sum_{\b} R^{\a \b}_{\g \l }=0,
\ee
these two Hamiltonians give rise to the same $\A$ and $ \bar \A$.
Regarding one--point functions $\langle n^{\a}_i \rangle $,
these models are the same. So,
we call these models, regarding one--point functions, equivalent.

However, $\A$ and $\bar \A$ do not determine $B$, $C$, $\bar B$,
and $\bar D$ uniquely. The stochastic condition
\be
\sum_{\a \b } H^{\a \b }_{\g \l }=0,
\ee
results in some constraints on $B$, $C$, $\bar B$,
and $\bar D$:
\bea
&&\sum_{\a }(C^\a_\b -B^\a_\g )=0\cr
&&\sum_{\a }(-\bar B^\a_\b +\bar D^\a_\g )=0,
\eea
So, the sum of all elements of any column of $B$ ($C$) should be the same
\be \label{cc}
\sum_{\a }C^\a_\b =\sum_{\a } B^\a_\b =f.
\ee
Then, the state $\langle   s\vert $ is the left eigenvector of $B$ and $C$, with
the same eigenvalue $f$. $\bar B$ and $\bar D$ have also
the same property, of course with different eigenvalue $g$.

Changing $B$ and $C$ according to the {\it gauge} transformation,
\bea \label{gg}
C^\a_\b \to  {C'}^\a_\b = C^\a_\b -f^{\a}&
 {\rm or}& C'=C-\vert f\rangle \langle   s\vert \cr
B^\a_\b \to  {B'}^\a_\b = B^\a_\b -f^{\a}&
{\rm or}& B'=B-\vert f\rangle \langle   s\vert
\eea
does not change $\A$ .
With a suitable choice of $f^{\a}$:
\be
\sum _{\a }f^{\a}=f,
\ee
the sum of the elements of any column of $B$ or $C$ can be set to zero.
In this gauge, the eigenvalues of $B$ and $C$ for the eigenvector
$\langle   s\vert $ will be zero.

\section{One-point functions}
\noindent To solve (\ref{ndot2}), we introduce the vector $\N_k$
\be
\N_k :=\pmatrix{\langle n^{1}_k \rangle \cr \langle n^2_k \rangle \cr
\cdot \cr \cdot \cr \cdot \cr \langle n^N_k \rangle }.
\ee
Equation (\ref{ndot2}) can then be written as
\be\label{Nk}
\dot \N_k=-(B+\bar B)\N_k +C \N_{k-1}+\bar D \N_{k+1}.
\ee
Introducing the generating function $G(z,t)$,
\be \label{G}
G(z,t)=\sum_{-\infty}^{\infty} \N_k(t) z^k,
\ee
one arrives at,
\be
\dot G(z,t)=\big[-( B+\bar B )+z\ C +z^{-1}\ \bar D\big] G(z,t),
\ee
the solution to which is
\be
G(z,t)=\exp \big( t[-( B+\bar B )+z\ C +z^{-1}\ \bar D]\big) G(z,0).
\ee
$\N_k(t)$'s are the coefficients of the Laurent expansion of $G(z,t)$, so
\be \label{gen}
\N_k(t)={1\over 2\pi i} \sum_{m=-\infty}^{\infty} \oint \d z \ z^{m-k-1}
\exp \big( t[ -(B+\bar B )+z\ C +z^{-1}\ \bar D]\big) \N_m(0).
\ee
This is the formal solution of the problem, which is of the form
\be
\N_k(t)=\sum_m \G_{km}(t) \N_m(0).
\ee

\subsection{Some special cases}
\noindent We now consider special choices for $B$, $C$, $\bar B$,
and $\bar D$.

\subsubsection{The matrices $B$, $C$, $\bar B$, and $\bar D$ are
two--dimensional (the single--species case)}

We can use the gauge transformation to make $\langle s|$ the simultaneous
null left--eigenvector of $B$, $C$, $\bar B$, and $\bar D$. In this
gauge, one
has
\bea\label{2DA}
B&=&|u\rangle\langle b|,\cr
C&=&|u\rangle\langle c|,\cr
\bar B&=&|u\rangle\langle\bar b|,\cr
D&=&|u\rangle\langle d|,
\eea
where
\be
|u\rangle :=\pmatrix{1\cr -1\cr}.
\ee
This means that it is orthogonal to $\langle s|$, and is a simultaneous
right--eigenvector of $B$, $C$, $\bar B$, and $\bar D$. Using (\ref{2DA}),
one can easily calculate the exponential in (\ref{gen}):
\be
\exp \big( t[ -(B+\bar B )+z\ C +z^{-1}\ \bar D]\big) =1+{{e^{tg(z)}-1}\over
{g(z)}}|u\rangle\langle g(z)|,
\ee
where
\be
\langle g(z)|:=-\langle b|-\langle\bar b|+z\langle c|+z^{-1}\langle\bar d|,
\ee
and
\be
g(z):=\langle g(z)|u\rangle .
\ee
Now take $\langle v|$ and $|w\rangle$ to be the left--eigenvector of
$-B-\bar B+C+\bar D$ dual to $|u\rangle$, and the right--eigenvector of
$-B-\bar B+C+\bar D$ dual to $\langle s|$, respectively. One can normalize
these, so that
\bea
\langle v|u\rangle &=& 1,\cr
\langle s|w\rangle &=& 1.
\eea
Of course, $\langle v|$ is orthogonal to $|w\rangle$. Then,
\be
\exp \big( t[ -(B+\bar B )+z\ C +z^{-1}\ \bar D]\big) =
e^{tg(z)}|u\rangle\langle v| +|w\rangle\langle s|+\langle g(z)|w\rangle
{{e^{tg(z)}-1}\over{g(z)}}|u\rangle\langle s|.
\ee
Acting this on $\N_m(0)$, and noting that
\be
\langle s|N_m(0)=1,
\ee
it is seen that
\bea
N_k(t)&=&|w\rangle\langle s|N_k(0) +{1\over 2\pi i} \sum_{m=-\infty}^{\infty}
\oint \d z \ z^{m-k-1}e^{tg(z)}|u\rangle\langle v|\N_m(0)\cr
&=&|w\rangle +{1\over 2\pi i} \sum_{m=-\infty}^{\infty}
\oint \d z \ z^{m-k-1}e^{tg(z)}|u\rangle\langle v|\N_m(0),
\eea
or
\be
\langle v|N_k(t)={1\over 2\pi i} \sum_{m=-\infty}^{\infty}\oint \d z \
z^{m-k-1}e^{tg(z)}|u\rangle\langle v|\N_m(0).
\ee
This is equivalent to (\ref{2D}).

\subsubsection{$C= pB, \qquad \bar D=q \bar B $}

 Using (\ref{sbc})
\be
(1-p)\langle   s\vert B=(1-q)\langle   s\vert \bar B =0.
\ee
means that $p=1$ or $\langle   s\vert B=0$,
and $q=1$ or $\langle   s\vert \bar B=0$.
If $\langle   s\vert$ is not the left null eigenvector of $B$ and
$\bar B$, then $p=q=1$. So, we will have $B=C$, and
$\bar D=\bar B$.
Now, using the definition of $\A $
\bea
&& \A^\a_{ \b \g}=C^\a_\b - C^\a_\g =\sum_{\l} H^{\l \a}_{\b \g}\cr
&& \A^\a_{\g \b}=C^\a_\g - C^\a_\b =\sum_{\l} H^{\l \a}_{\g \b}.
\eea
For $\a\ne \b$, and $\a \ne \g$,  all the terms in the
right hand side summations in the above equations are reaction rates and
should be non--negative, but the sum of the left hand sides is zero. So,
\be
C^\a_\b = C^\a_\g =f^\a , \qquad {\rm for}\quad \g \ne \a\ne \b  .
\ee
All the elements of each row except the diagonal elements of $C$
( or $B$ ) are the same. That is,
\be
C=|f\rangle\langle s|+C',
\ee
where $C'$ is some diagonal matrix. The fact that $|s\rangle$ is a
left--eigenvector of $C$, shows that it should be a left--eigenvector of
$C'$ as well. And this demands $C'$ to be proportional to the
unit matrix. One can do the same arguments for $\bar B$ and $\bar D$.
So, after gauge transformation,
\be
C=B=u {\bf 1},\qquad \bar D=\bar B=v{\bf 1}.
\ee
Although, the time evolution of average densities can be written
in terms of $B$, $C$, $\bar B$, and $\bar D$ , the Hamiltonian $H$
is not uniquely be determined by these matrices.
There exist different Hamiltonians which are
equivalent, regarding one--point functions.
\bea
&&\sum_{\l} H^{\l \a}_{\b \g }=\A^\a_{\b \g }=u( \delta^\a_\b -
\delta^\a_\g )\cr
&&\sum_{\l} H^{\a \l}_{ \b \g }=\bar \A^\a_{ \b \g }=
v( \delta^\a_\g -\delta^\a_\b ).
\eea
All the elements of the $\b \b $ column of $H$ are zero. For $\a\ne\b$, the
elements of $H$ satisfy
\bea
&&\sum_{\l\ne\a ,\b}H^{\l\a}_{\a\b}+H^{\a \a}_{\a\b}+H^{\b \a}_{\a\b}=u\cr
&&\sum_{\l\ne\a ,\b}H^{\l\b}_{\a\b}+H^{\a \b}_{\a\b}+H^{\b \b}_{\a\b}=-u\cr
&&\sum_{\l\ne\a ,\b}H^{\b\l}_{\a\b}+H^{\b \a}_{\a\b}+H^{\b \b}_{\a\b}=v\cr
&&\sum_{\l\ne\a ,\b}H^{\a\l}_{\a\b}+H^{\a \a}_{\a\b}+H^{\a \b}_{\a\b}=-v.
\eea
In general, these sets of equations have several solutions, but for the
one--species case, the reaction rates are the following

\be
A\p  \to \cases{\p A & $\qquad \L_{12}$ \cr
A A & $\qquad u - \L_{12}$\cr
\p\p & $\qquad v- \L_{12}$\cr}
\ee
\be
\p A  \to \cases{A\p  & $\qquad \L_{21}$ \cr
A A & $\qquad v - \L_{21}$\cr
\p\p & $\qquad u- \L_{21}$\cr}
\ee
The above system, with no diffusion, has been studied in \cite{AM}.
There, the $n$--point functions have been investigated.
This solution can be generalized to the multispecies case. For $\a \ne \b$
\bea
A_\a A_\b \to A_\b A_a, &&\qquad \L_{\a\b}\quad \a ,\b =1\cdots N \cr
A_\a A_\b  \to A_\a A_\a, &&\qquad u-\L_{\a\b}\cr
A_\a A_\b \to A_\b A_\b, &&\qquad v-\L_{\a\b}.
\eea
The only constraint is the non--negativeness of the reaction rates:
\be
u\geq \L_{\a\b}\geq 0\qquad v\geq \L_{\a\b}\geq 0.
\ee
This model has $N(N-1)+2$ free parameters. However, only the two
parameters $u$ and $v$ appear in the time evolution equation of average
densities:
\be \label{ndot3}
\langle \dot n^{\a}_i \rangle = -(u+v)\langle  n^{\a}_{i}\rangle
+u\langle  n^{\a}_{i-1}\rangle +v\langle  n^{\a}_{i+1}\rangle  .
\ee
As it is seen, dynamics of average densities of different particles
decouple, and despite the complex interactions of the model,
$\langle \dot n^{\a}_i \rangle$'s can be easily calculated.
But in the time evolution of two-point functions  $\L_{\a\b}$'s  appear
as well. So, although models with different exchanging rates ($\L_{\a\b}$)
and same initial conditions have the same average densities, their
two--point functions generally are not the same.

\subsubsection{$B,\ \bar B,\ C, \ \bar D $ commute}

Generally, the gauge transformation do not preserve the commutation relation
of $B$ and $C$ ( and that of $\bar B$ and $\bar D$ ). But if $B$ and $C$
commute, there is a gauge transformation which leaves the transformed $B$
and $C$ commuting. If we choose $\vert f\rangle $ to be a right eigenvector
of $B$ and $C$ dual to $\langle s\vert $, that is
\be
B\vert f\rangle =C\vert f\rangle =f\vert f\rangle ,
\ee
then $B':=B-|f\rangle\langle s|$ and $C':=C-|f\rangle\langle s|$ commute. If
\be
\langle s\vert f\rangle =f,
\ee
then $\langle s\vert$ times $B'$ and $C'$ will be zero.
So, if $B$, $C$, $\bar B$, and $\bar D$ commute with each other, there
exists a suitable gauge transformation that makes their eigenvalue
corresponding to $\langle s|$ zero, while they remain commuting:
\be
\langle   s\vert  B=\langle   s\vert  C=\langle   s\vert  \bar B=
\langle   s\vert  \bar D=0,
\ee
Denote, the matrix which simultaneously diagonalize these four matrices by
$U$, diagonalized matrices by primes, and their eigenvalues by
$b^{\a}$, $c^{\a}$, $\bar b^{\a}$, and $\bar d^{\a}$, respectively. We have
\bea
&& \langle \O \vert  B'=\langle \O \vert  C'=\langle \O \vert  \bar B'=
\langle \O \vert  \bar D'=0\cr
&& \langle \O \vert =\langle s \vert U.
\eea
We take $b^N=c^N=\bar b^N=\bar d^N=0$, and normalize $\langle \O\vert$ and
$U$ so that
\be
\langle \O \vert =(0\ 0\ \cdots \ 0\ 0),
\ee
and
\be
\sum_{\a} U_{\a \b}=\delta_{N\b}.
\ee
$U$ will also diagonalize the exponential in (\ref{gen}). So we have
\be \label{sp}
\N'_k(t)={1\over 2\pi i} \sum_{m=-\infty}^{\infty} \oint \d z \ z^{m-k-1}
\exp \big( t[ -B'-\bar B' +z\ C' +z^{-1}\ \bar D']\big) \N'_m(0),
\ee
where
\be \label{n'}
\N'_k(t):=U^{-1}\N_k(t).
\ee
The matrix in the argument of the exponential in (\ref{sp}) is
diagonal, so the integral can be easily calculated:
\be \label{int}
{\cal I}:={1\over 2\pi i} \oint \d z \ z^{m-k-1}
\exp \big( t[- b^\a -\bar b^\a +z\ c^\a +z^{-1}\ \bar d^\a ]\big).
\ee
Introducing $w:=\sqrt{\bar d^\a \over c^\a}z$, one arrives at
\be \label{int2}
{\cal I}:=({\bar d^\a \over c^\a})^{m-k\over 2}
{e^{-t (b^\a  +\bar b^\a )}\over 2\pi i} \oint \d w \ w^{m-k-1}
\exp \big( \sqrt{c^\a \bar d^\a }t(w +w^{-1})\big),
\ee
which can be written in terms of modified Bessel functions
\be \label{int3}
{\cal I}:=({\bar d^\a \over c^\a})^{m-k\over 2}
e^{-t (b^\a  +\bar b^\a )} {\rm I}_{k-m}(2\sqrt{c^\a \bar d^\a }t).
\ee
Then,
\be \label{nkm}
\N_k(t)=\sum_{m=-\infty}^{\infty}U{\rm diag}\left\{ ({\bar d^\b \over c^\b})^
{m-k\over 2}e^{-t (b^\b  +\bar b^\b )}{\rm I}_{k-m}(2\sqrt{c^\b\bar d^\b}t)
\right\} U^{-1}\N_m (0).
\ee
Note that the right--hand side of (\ref{int3}) is $\delta_{k,m}$ for
$\a =N$, since the $N$-th eigenvalue of $B$, $C$, $\bar B$, and $\bar D$ is
zero.

One can start with four special diagonal matrices, and then construct the
Hamiltonians with different reaction--diffusion rates. Not all diagonal
matrices lead to physical stochastic models: negative reaction rates may be
obtained. Considering the large time behaviour of average number densities,
one can show that
\be
|{\rm Re}(\sqrt{ c^{\a} \bar d^{\a}})|\leq{\rm Re}(b^{\a}+\bar b^{\a}),
\ee
which also shows that
\be \label{bp}
{\rm Re}(b^{\a}+\bar b^{\a})\geq 0.
\ee

Now,  we consider a special choice for $U$:
\be
U^\a_\b=\delta^\a _N-(1-\delta^\a _N)\delta^\a _\b.
\ee
Then
\bea
B^\a_\b&=& b_{\b}(\delta^\a _\b -\delta^\a _N)\cr
C^\a_\b&=& c_{\b}(\delta^\a _\b -\delta^\a _N)\cr
\bar B^\a_\b&=& \bar b_{\b}(\delta^\a _\b -\delta^\a _N)\cr
\bar D^\a_\b&=& \bar d_{\b}(\delta^\a _\b -\delta^\a _N).
\eea
Now, consider
\be
\A^\a_{\b\g}=\sum_{\l}H^{\l\a}_{\b\g}=-b^{\a}\delta^\a _\g+
c^{\a}\delta^\a_\b +(-b_{\g}+c_{\b} )\delta^\a_N.
\ee
For $\a\ne\g$ and $\a\ne\b$,
\be
\sum_{\l}H^{\l\a}_{\b\g}\geq 0\qquad \sum_{\l}H^{\l\a}_{\g\b}\geq 0.
\ee
So, taking $\b ,\g\ne N$ and $\a =N$,
\be
b^{\g}\geq c^{\b}.
\ee
The same reasoning is true for $\bar b^{\g}$ and $\bar d^{\b}$:
\be
\bar b^{\g}\geq \bar d^{\b}.
\ee
Here too, similar to the previous example, the above choices for $B$, $C$,
$\bar B$, and $\bar D$ do not determine $H$ uniquely. One particular
solution for the reaction rates is

For $\a \ne N$

\be
A_N A_\a  \to \cases{A_\a A_N, & $\qquad \L_{N\a}$ \cr
A_\a A_\a, & $\qquad \bar d_\a - \L_{N\a}$\cr
A_N  A_N, & $\qquad b_\a - \L_{N\a}$\cr}
\ee
\be
A_\a A_N  \to \cases{ A_N A_\a, & $\qquad \L_{\a N}$\cr
A_\a A_\a, &$\qquad c_\a - \L_{\a N}$\cr
A_N A_N, &$\qquad \bar b_\a - \L_{\a N}$\cr}
\ee

and, for $\a ,\b\ne N$
\be
A_\a A_\b \to \cases{ A_\a A_N, &$ \qquad  b_\b -c_\a - \L_{\a\b}$\cr
A_N A_\b, &$\qquad \bar b_\a -\bar d_\b - \L_{\a\b}$ \cr
A_N A_N, &$ \qquad  \L_{\a\b}$\cr}
\ee

For $\a \ne \b $, the following reactions may also occur.
For $\a < \b $

\be
A_\a A_\b \to\cases{ A_\b A_\a, &$\qquad c_\a $\cr
                     A_\b  A_\b, &$\qquad -c_\a+\bar d_\b  $\cr}
\ee
and for $\a >\b $
\be
A_\a A_\b \to\cases{ A_\b A_\a, &$\qquad \bar d_\b$ \cr
                     A_\a  A_\a, &$\qquad c_\a -\bar d_\b $  \cr}
\ee
The constraint of non--negativeness of the reaction rates leads to
\bea
c_\a\leq d_\b\leq c_\g &\qquad &\a <\b <\g\cr
0 \leq\L_{\a\b}\leq  b_\b -c_\a&&\cr
\L_{\a\b}\leq \bar b_\a -\bar d_\b &&\cr
0 \leq \L_{N \a }\leq \bar d_a &&\cr
0 \leq \L_{N \a }\leq b_a &&\cr
0 \leq \L_{\a N}\leq \bar b_a &&\cr
0 \leq \L_{\a N }\leq c_a &&
\eea
\subsubsection{type--change invariance}

Suppose $B$, $C$, $\bar B$, and $\bar D$ have the property
\be
B^{\a +\g}_{\b +\g}=B^\a_\b,
\ee
and the same for the other three matrices. Note that the indices of these
matrices are defined periodically, so that $N+\a$, as an index, is
equivalent to $\a$. This is in fact a special case of commuting matrices
discussed earlier. One can use (\ref{nkm}). To do so, one should know the
simultaneous eigenvectors of $B$, $C$, $\bar B$, and $\bar D$, and their
corresponding eigenvalues. It is not difficult to see that the eigenvectors
are
\be
U^\a_\b ={1\over{\sqrt{N}}}\exp\left({{i2\pi\a\b}\over N}\right).
\ee
The corresponding eigenvectors of $B$, for example, are
\be
b^\b=\sum_\a B^\a_0\exp\left(-{{i2\pi\a\b}\over N}\right).
\ee
Finally, the matrix elements of the inverse of $U$ are
\be
(U^{-1})^\a_\b ={1\over{\sqrt{N}}}\exp\left(-{{i2\pi\a\b}\over N}\right).
\ee
These can be put directly in (\ref{nkm}).

\section{Large time behaviour of average densities}
The large--time behaviour of the system is deduced through a
steepest--descent analysis of the formal solution (\ref{gen}). One should
consider the eigenvalues and the eigenvectors of the $z$--dependent matrix
\be
M(z):=-(B+\bar B)+zC+z^{-1}\bar D.
\ee
Denote the eigenvalues of this matrix by $\l^\a(z)$. As for any value of
$z$, the matrix $M$ has $\langle s|$ as its left eigenvector corresponding
to the eigenvalue zero, $M$ will have a right eigenvector $|w\rangle$ dual
to $\langle s|$. $|w\rangle$ is $z$--dependent, but one can normalize it so
that
\be
\langle s|w(z)\rangle =1.
\ee
The fact that $\N$ should not blow up at $t\to\infty$ assures that the
real--part of the eigenvalues of $M(z)$ are non--positive (at least for
$|z|=1$). If all of the other eigenvalues have negative real--parts, then
at $t\to\infty$ only $|w\rangle$ survives. That is,
\bea
\N_k(\infty)&=&{1\over{2\pi i}}\sum_m\oint\rd z\; z^{m-k-1}|w(z)\rangle
\langle s|\N_m(0)\cr
&=&|w(1)\rangle.
\eea
We have used $\langle s|\N_m(0)=1$. This could also be obtained directly,
using the evolution equation (\ref{ndot2}), by setting $\dot\N_k$ equal to
zero and assuming $\N_k$ independent of $k$. So, the final state of the
system is the eigenvector of $-(B+\bar B)+C+\bar D$, corresponding to the
eigenvalue zero.

To investigate the next--to--leading term at $t\to\infty$, consider the
other eigenvalues of $M(z)$. Suppose that at $z=z_0^\a$, $\l^\a$ is
stationary. There may be more than one point having this property. So, we
will have a set consisting of $z_{0a}^\a$'s. Each of these points
corresponds to a stationary eigenvalue $\l_{0a}^\a$. We choose that
$z_{0a}^\a$ the corresponding eigenvalue of which has the largest
real--part. Denote this point by $z_0$, its corresponding stationary
eigenvalue by $\l_0$, and its corresponding right eigenvector by
$|v_0\rangle$. The next--to--leading term in $\N$ is then
\be
\N_k^{(1)}\sim z_0^{-k}e^{t\l_0}|v_0\rangle .
\ee
Note that $z_0$ is not necessarily a phase, its modulus may be different
from 1.

\newpage


\begin{thebibliography}{99}
\bibitem{ADHR} F. C. Alcaraz, M. Droz, M. Henkel, \& V. Rittenberg;
               Ann. Phys. {\bf 230} (1994) 250.
\bibitem{KPWH} K. Krebs, M. P. Pfannmuller, B. Wehefritz, \&
               H. Hinrichsen; J. Stat. Phys. {\bf 78}[FS] (1995) 1429.
\bibitem{HS} H. Simon; J. Phys. {\bf A28} (1995) 6585.
\bibitem{PCG} V. Privman, A. M. R. Cadilhe, \& M. L. Glasser; J. Stat.
               Phys. {\bf 81} (1995) 881.
\bibitem{HOS1} M. Henkel, E. Orlandini, \& G. M. Sch\"utz; J. Phys. {\bf A28}
               (1995) 6335.
\bibitem{HOS2} M. Henkel, E. Orlandini, \& J. Santos; Ann. of Phys. {\bf 259}
               (1997) 163.
\bibitem{AL} A. A. Lusknikov; Sov. Phys. JETP {\bf 64} (1986) 811.
\bibitem{AKK} M. Alimohammadi, V. Karimipour, M. Khorrami; Phys. Rev.
              {\bf E57} (1998) 179.
\bibitem{RK} F. Roshani, M. Khorrami; Phys. Rev. {\bf E60} (1999) 3393.
\bibitem{AKK2} M. Alimohammadi, V. Karimipour, M. Khorrami; J. Stat.
               Phys. {\bf 97} (1999) 373.
\bibitem{SBH} R. Schoonover,D. Ben--Avraham,S. Havlin, R. Kopelman, \&
              G. Weiss;  Physica {\bf A171} (1991) 232.
\bibitem{TW}  D. Toussaint, F. Wilczek; J. Chem. Phys. {\bf 78} (1983) 2642.
\bibitem{KR} K. Kong, S. Render; Phys. Rev. Lett. {\bf 52} (1984) 955.
\bibitem{BPZ} D. Ben--Avraham, V. Privman,D. Zhong; Phys. Rev. {\bf E52}
              (1995) 6889.
\bibitem{IKR} I. Ispolatov, P. L. Krapivsky,S. Render; Phys. Rev. {\bf E52}
             (1995) 25406889.
\bibitem{LP} J. W.Lee, V. Privman; J. Phys. {\bf A30} (1997) L317.
\bibitem{ZK} Z. Koza; cond-mat/9601072.
\bibitem{VK}  V. Karimipour; Phys. Rev.  {\bf E59} (1999) 205.
\bibitem{KC}  K. Kim, K. H. Chang; J. Phys. Soc. Jpn. {\bf 68} (1999) 1450.
\bibitem{AA}  M. Alimohammadi, N. Ahmadi; cond-mat/0002265 , to appear in
              Phys. Rev. E (2000).
\bibitem{OM}  G. Odor, N. Menyhard; cond-mat/0002199.
\bibitem{FJ}  F. H. Jafarpour; cond-mat/0004357.
\bibitem{GS}  G. M. Sch\"utz; J. Stat. Phys. {\bf 79}(1995) 243.
\bibitem{AM}  A. Aghamohammadi, M. Khorrami; cond-mat/0005532
\end{thebibliography}
\end{document}